\documentclass{article}
\usepackage[utf8]{inputenc}
\usepackage{authblk}
\usepackage{graphicx} 
\usepackage{amsmath}
\usepackage{amsfonts}
\usepackage{amssymb}
\usepackage{xcolor}
\usepackage{array}

\usepackage{pdflscape}
\usepackage{rotating}
 \usepackage{lineno}

\usepackage{geometry}
\usepackage{tikz}
\usepackage{ulem}

\usepackage[backend=biber, style=numeric-comp, sorting=none]{biblatex}
\AtEveryBibitem{\clearfield{month}}
\AtEveryCitekey{\clearfield{month}}
\renewbibmacro{in:}{}
\usepackage{etoolbox}
\AtBeginBibliography{%
  \sloppy
  \emergencystretch=3em
}

\makeatletter
\AtEveryCitekey{%
  \ifcsundef{blx@entry@refsegment@\the\c@refsection @\thefield{entrykey}}
    {\csnumgdef{blx@entry@refsegment@\the\c@refsection @\thefield{entrykey}}
      {\the\c@refsegment}}
    {}%
}

\defbibcheck{onlynew}{%
  \ifnumless
    {0\csuse{blx@entry@refsegment@\the\c@refsection @\thefield{entrykey}}}
    {\the\c@refsegment}
    {\skipentry}
    {}%
}
\makeatother

\newcommand*\circled[1]{\tikz[baseline=(char.base)]{
            \node[shape=circle,draw,inner sep=2pt] (char) {#1};}}

\title{Hydrodynamics of microalgae in sea spray}

\author[a]{Elettra Figà Talamanca}
\author[a,\dag,*]{Juliette Pierre}
\author[b]{Eva Ternon}
\author[b]{Paula Cohen}
\author[c]{Maarten De Rijcke}
\author[a,\dag,*]{Hélène de Maleprade}

\affil[a]{{\footnotesize Institut Jean Le Rond d'Alembert, Sorbonne Université, CNRS, 75005 Paris, France}}
\affil[b]{{\footnotesize Sorbonne Université, CNRS, Laboratoire d’Océanographie de Villefranche (UMR 7093), 06230 Villefranche-sur-mer, France}}
\affil[c]{{\footnotesize Flanders Marine Institute (VLIZ), InnovOcean Campus, Jacobsenstraat 1, 8400 Oostende, Belgium}}

\date{\today}

\begin{document}
\maketitle

\textsuperscript{\dag}J.P. and H.M. contributed equally to this work.

\textsuperscript{*}E-mail: juliette.pierre@cnrs.fr, helene.de\_maleprade@sorbonne-universite.fr\\

Harmful algal bloom $|$ Sea spray $|$ Microalgae $|$ Bubble bursting $|$ Jet drop\\

\begin{refsegment}
\begin{abstract}
Airborne transmission of biological matter is likely to cause numerous respiratory distress in coastal populations during certain harmful algal blooms \cite{Berdalet2022}. While we know that dissolved algal toxins aerosolize \cite{Ternon2024}, the contribution of particle-associated toxins is still unknown. The observation of microalgae in the atmosphere \cite{Veron2015, Alsante2021} suggests that cells or cell debris could be a carrier. Exchange of particulate matter between ocean and atmosphere is driven by sea spray \cite{Deike2022}: microplastics and bacteria aerosolize in jet droplets, following the bursting of air bubbles \cite{Blanchard1989, Shaw2023}.
However, current models for particle capture only consider surface effects \cite{Dubitsky2023, Ji2022}, which fail to explain how large microalgal cells (5-50 µm) can be entrapped in jet drops (100-800 µm).
Here, we demonstrate with a variety of species that microalgae are consistently captured in jet drops, and derive a new particle capture model based on the hydrodynamics of the jet. Using the enrichment factor, which measures the increase in concentration in the first jet drop relative to the bulk, we uncover two regimes. The transition between regimes strongly depends on cell size. In addition, we highlight the dynamics of particles within the jet and derive an enrichment model based on hydrodynamics rather than surface considerations. This work widens and generalizes established results on bacterial and microplastic entrapment in jet drops, resolving discrepancies found in the literature. 
Finally, we demonstrate that sea spray can aerosolize toxic \textit{Ostreopsis} cf. \textit{ovata} cells. This reveals a potential particulate pathway for airborne toxicity during harmful algal blooms and raises the question of the relative contribution of particles to that of dissolved toxins. 
\end{abstract}


Airborne toxicity during harmful algal blooms is a growing public health issue, as there is an increasing body of reports of respiratory irritations in coastal populations \cite{Illoul2012, Berdalet2022}. To date, a few dinoflagellate species are related to respiratory irritation \cite{Fleming2011, Ternon2025, Murray2026}, including \textit{Ostreopsis} cf. \textit{ovata} (OO) \cite{Lim2023, Casabianca2014}. This species has been blooming along the Mediterranean coasts for over 20 years, and has recently established itself along the Atlantic coast, with a first documented outbreak in 2021 in the Bay of Biscay \cite{Chomerat2022}. During blooms, OO cells produce potent phycotoxins at a level of pg per cell \cite{Honsell2011} that are soluble in seawater though rapidly degraded by sunlight and bacterial activity \cite{Ternon2022}. 

Although airborne toxicity is well-established for \textit{Ostreopsis} harmful algal blooms \cite{Lim2023, Casabianca2014, Chomerat2022, Honsell2011, Ternon2022}, mechanisms underlying the transfer of toxins are still under debate. Airborne phycotoxins are the most probable cause of toxicity \cite{Ternon2024}, and their concentration in sea spray aerosols correlates with particulate phycotoxins in seawater. The concentration of particulate phycotoxins in the water column and the cellular toxin content both increase with cell concentration over the duration of a bloom \cite{Medina-Perez2021, Gemin2020}.
Yet, given that OO phycotoxins are hydrophilic and unstable in dissolved form \cite{Ternon2022}, airborne toxicity may also be caused by particle-associated toxins, via the expulsion of cell fragments and intact OO cells.

\begin{figure*}
\centering
\includegraphics[width=1\linewidth]{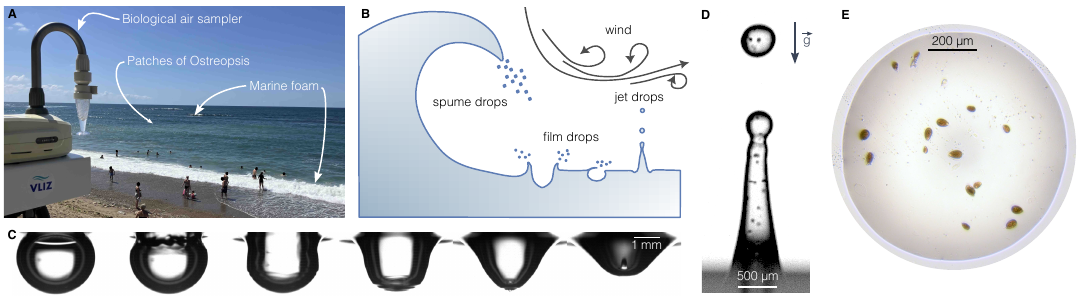}
\caption{Detection of \textit{Ostreopsis} cf. \textit{ovata} in the field and in the laboratory. \textbf{A}: Field experiment. A biological air sampler stands $20$~m away from the shoreline. Marine foam and OO patches are visible in the picture. \textbf{B}: Schematic representation of the three main mechanisms of sea spray production: spume drops, by wind-driven wave crest breakup, and film and jet drops, following the bursting of surface bubbles. \textbf{C}: Side view of an underwater temporal sequence of a bubble cavity collapse. The time interval between images is $\Delta t=0.48$~ms. \textbf{D}: Jet fragmentation after bubble bursting, for a bubble with radius $R_b=1.2$~mm, rising from $H=1.5$~cm, in a culture with $C_{bulk}=3.2 \times10^5$ cells/mL. Small dark points are OO cells. \textbf{E}: Microscopic view of the first, fastest jet drop in D, collected on a glass slide. The droplet contains several motile OO cells and cells fragments.}
\label{fig:1}
\end{figure*}

Consistent with reports of microorganisms in the atmosphere \cite{Alsante2021, Ciminiello2014}, we captured OO cells in the air during an \textit{Ostreopsis} bloom in 2025 in the Basque Country. We installed a bioaerosol collector along the shore at high tide to collect the ambient air (see Fig.~\ref{fig:1}A and Methods). Over three days of measurements, we systematically captured numerous intact and ruptured OO cells in the collector (see Supplementary Information (SI)). Specifically, we measured a flux of 10 to 20 cells per hour, potentially inhaled by beach-goers, sea-side workers and inhabitants. This key observation strengthens the hypothesis that particulate matter, such as cell fragments and intact cells, could contribute to airborne toxicity \cite{Ternon2024}. Motivated by this field observation, we developed laboratory experiments to explore one of the mechanisms underlying the ejection of microalgae into the air.

Sea spray and marine aerosols are the main driver of ocean--atmosphere exchanges \cite{Deike2022} and mediate transfer of biological and non-biological matter, influencing climate, biogeochemical cycles, and air quality \cite{Alsante2021, Blanchard1989} through several mechanisms illustrated in Fig.~\ref{fig:1}B. Wave crests break up under high wind stress and eject spume drops, while droplets impacting the sea and surface bubble bursting emit film and jet drops. Their typical size ranges from micrometers to several hundred micrometers~\cite{Deike2022, Veron2015}. Once airborne, droplets can be propelled over vast distances by wind. They subsequently evaporate and shrink, which enables long-range atmospheric transport~\cite{Veron2015}.

Here, we focus on jet drops that are larger than film droplets and therefore more likely to contain particles as large as OO cells, which are about 35~µm \cite{Honsell2013}. The emission of jet drops during the bursting of surface bubbles is a widely investigated phenomenon whose underlying physical mechanisms are now well understood for surfactant-free solutions \cite{Lai2018, Blanco-Rodriguez2020, Ganan-Calvo2021, Berny2020, Berny2022, Gordillo2023}. When a bubble bursts at the surface of a liquid, the thin free film that caps the bubble ruptures and ejects microdroplets (film drops). The resulting open cavity relaxes within milliseconds as capillary waves propagate along the liquid-gas interface toward the bubble’s base (see Fig.~\ref{fig:1}C and Supplementary Movie 1). When they converge, the cavity reverses and ejects a high-velocity Worthington jet upwards. Due to the Rayleigh-Plateau instability, this jet ultimately fragments into droplets, known as jet drops (Fig.~\ref{fig:1}D). In surfactant-free solutions, the size and velocity of ejected droplets directly correlate with the initial bubble radius \cite{Ganan-Calvo2017}.

Since the pioneering work of Blanchard in the 1970s on water-to-air bacterial transport \cite{Blanchard1970}, few studies have focused on the transfer of contaminants—whether chemical \cite{Neel2022, Pierre2022, Pico2024}, biological \cite{Bezdek1972, Blanchard1972}, or particulate \cite{Sakai1988, Ji2022, Dubitsky2023, Shaw2023, Oehlschlagel2024, Chu2025, Catarino2023}—from the sea to the atmosphere. These works collectively demonstrate that the primary jet drop entraps bacteria and particles suspended in water. Notably, all report that particle or cell concentration in droplets is consistently larger than in the water column. This effect is quantified by the enrichment factor $E$, which is the ratio of droplet to bulk concentration ($E \equiv C_{drop}/C_{bulk}$) and can strikingly reach several hundreds.

Enrichment factors larger than one are puzzling. To date, enrichment is modelled as a surface effect: particles scavenged at the interface of rising bubbles end up in the jet \cite{MacIntyre1972, Blanchard1989, Ji2022, Dubitsky2023}. While the enrichment dependence on bulk concentration has not yet been explored, several experimental studies report an increase of $E$ for smaller drop radii $r_d$ \cite{Dubitsky2023, Blanchard1970, Bezdek1972}, and an increase of $E$ with the distance $H$ travelled through the water column \cite{Blanchard1981, Ji2022, Shaw2023}. Yet, the current surface model fails to explain the entrapment of large particles, such as OO cells. In addition to our report of OO cells in the air, recent field studies highlight the presence of a variety of microalgae and diatom cells in sea spray \cite{Sinnett2025, Alsante2021, Tesson2016}. These microorganisms span a broad size range, from a few to several tens of micrometers, and their concentration strongly varies in time and space. 

Here, we explore how microalgae are captured in sea spray with an analogue lab experiment, considering the first droplet ejected by the bursting of a single bubble at the surface of an algal bath. In Fig.~\ref{fig:1}D and E, we confirm field observations as we show that sea spray can contain cells as large as \textit{Ostreopsis} cf. \textit{ovata}. This discovery motivated a physical exploration of the phenomenon in order to draw a generic and predictive law. For this purpose, we developed laboratory investigations based on a model non-toxic and well-characterised microalga, \textit{Chlamydomonas reinhardtii} (CR) \cite{Harris2009, Goldstein2015, Jeanneret2016}.
We then explore the first drop enrichment in CR as we extensively vary bubble characteristics (size and release height), cell concentration and microalgae properties. We identify two distinct regimes of enrichment depending on algal concentration, which we model using scaling arguments. 
Finally, we extend the generality of our work by including multiple microalgae with various sizes and shapes, to widen the ecological implications of our findings.

\section*{Algae in sea spray}
We investigate whether first drops carry wild-type CR. To develop a mechanistic physical model, we initially focus on CR, only exploring other species later. Our experimental setup described in the Methods allows us to explore the bursting of single bubbles: we follow the jet and drop dynamics and observe first drop contents under the microscope. We vary the bubble radius--hence $V_{drop}$ the drop volume--and cell concentration $C_{bulk}$ in the bulk. We validate that the presence of microalgae affects neither the size nor the dynamics of ejected droplets (see SI). Both remain related to the bubble size through the empirical laws established for homogeneous liquids~\cite{Ganan-Calvo2017, Lai2018}.

\subsection*{Number of cells}
\begin{figure}[h!]
\centering
\includegraphics[width=1\linewidth]{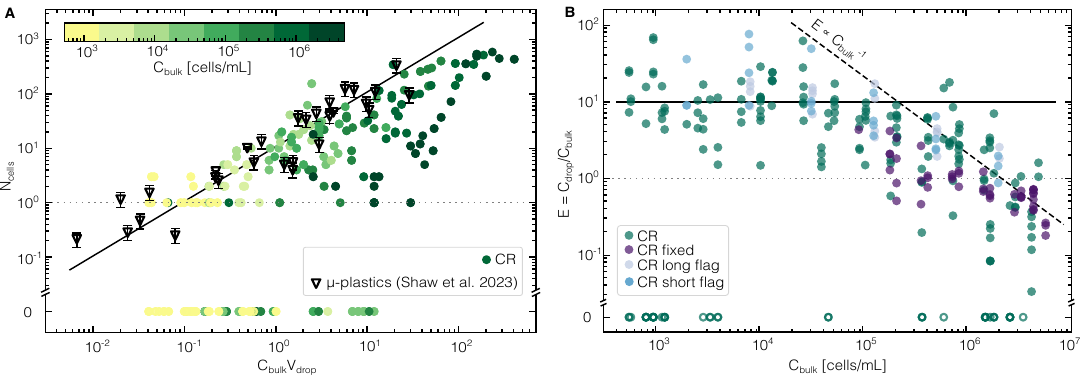}
\caption{First drop content.
\textbf{A:} Number of cells in the first droplet $N_{cells}$ as a function of the expected number of cells from a volume $V_{drop}$ taken in the bulk with concentration $C_{bulk}$, with constant $H\in[1.5, 2]$ cm. Dots show our data with wild-type CR. Color code stands for bulk concentration. Measurements with $N_{cells} = 0$ are reported at the bottom of the split logarithmic vertical axis. Grey dotted line marks the detection limit $N_{cells}=1$. Black triangles with the standard error and the black solid line show experiments with microplastics and the inferred model (Shaw et al. (2023)~\cite{Shaw2023}). 
\textbf{B:} Enrichment $E$ in the first drop as a function of the bulk concentration $C_{bulk}$ for all CR experiments at a fixed bubble rising distance $H \in[1.5,\,2]\,$cm. The black solid line marks $E=10$, the dashed line represents $E\propto C_{bulk}^{-1}$, the grey dotted line marks $E=1$. Dark green: wild-type CR, purple: fixed wild-type cells, light gray: long flagella CR mutant, blue: short flagella CR mutant. 
Measurements with $E = 0$ are reported at the bottom of the split logarithmic vertical axis. 
}
\label{fig:2}
\end{figure}

Fig.~\ref{fig:2}A shows the number of cells $N_{cells}$ in the first drop as a function of the expected cell number from an equivalent volume of bulk, $C_{bulk} V_{drop}$. We keep the rising distance of the bubble through the water column low and constant to $H\in[1.5,\;2]$~cm. Our results demonstrate that the first drop consistently captures CR as $N_{cells}$ varies over three orders of magnitude, up to a thousand. Note that we truncated the logarithmic y-scale to emphasize experiments with no cells. CR data points are color-coded from yellow to green, reflecting increasing $C_{bulk}$. 

We compare our results to those for microplastics of Shaw et al. (2023), in black triangles. The black line highlights their scaling law $N_{cells}=11 C_{bulk} V_{drop}$, equivalent to a constant $E=11$  \cite{Shaw2023}. While the microplastics results closely follow this scaling, our measurements only partially agree: if Shaw's model captures the upper trend, a non-negligible amount of data lie below. Specifically, our results deviate as the bulk concentration $C_{bulk}$ increases. The enrichment is surprisingly not constant but rather decreases at high concentrations. This apparent loss of efficiency has to our knowledge never been reported in the literature, and hints at an unexpected saturation in the enrichment process.

\subsection*{Algal concentration}
We unexpectedly uncovered that algal concentration dramatically impacts the enrichment (Fig.~\ref{fig:2}A), so we explored this dependency further. To begin, we verified that, although higher concentration implies more biomass, the liquid surface tension does not vary between experiments (see SI). Fig.~\ref{fig:2}B presents the enrichment factor $E$ as a function of the bulk algal concentration $C_{bulk}$, which we vary over 4 orders of magnitude, keeping the rising distance constant to $H\in[1.5,\;2]$~cm. The bubble size varies throughout the graph, but we demonstrate below and in the SI that $R_b$ has only a mild impact on $E$. 

At low concentrations, data with wild-type CR (dark green dots in Fig.~\ref{fig:2}B) show a non-negligible dispersion. Enrichment appears rather constant, about 10 (black horizontal line), remarkably close to Shaw's microplastics model with $E = 11$ \cite{Shaw2023}. Despite similar enrichment values, entrapment mechanisms may differ for microalgae as they are not expected to stick to the bubble interface. Therefore, the physical processes behind enrichment factors above unity remain unclear. 
At high concentrations ($C_{bulk} \geq 10^5$~cells/mL), dispersed data points initiate a marked decrease, following a trend close to $E\propto C_{bulk}^{-1}$, stressed by the dashed line, a behavior that had never yet been brought to light.

\subsection*{Bubble--algae interaction}
In order to alter the typical interaction distance between CR and the bubble surface, we use short- and long-flagella mutants (light gray and blue dots) and fixed wild-type CR cells (purple dots in Fig.~\ref{fig:2}B). This varies the geometrical extent of CR, as well as its hydrodynamic interactions to interfaces \cite{Kantsler2013}. These three additional sets of data surprisingly only strengthen our initial results with wild-type CR, as no clear difference emerges between all CR measurements. 

First, as fixed cells do not swim, we demonstrate that algal motility does not play a major role in the mechanism underlying encapsulation. Indeed, as CR swims at speeds about 100 µm/s and as a bubble bursts in a few milliseconds, CR travels over a distance of 0.1 µm at most. Additionally, varying flagellar length from 7 (short-) to 20~µm (long-flagella, see SI) does not substantially modify $E$. This suggests that either particle size does not impact the enrichment factor in contradiction to the literature \cite{MacIntyre1972}, or flagella do not impact the encapsulation phenomenon. As flagella length may modify the distance to the bubble surface \cite{Kantsler2013}, our unaffected enrichment factor hints that proximity to the bubble surface may not be key to large enrichment, diverging from past and current models. 

Finally, we verify in the SI that the enrichment increases with the rising distance $H$ of the bubble through the water column, keeping $C_{bulk} < 10^4$~cells/mL to be in the constant regime. This is consistent with microplastics \cite{Dubitsky2023, Shaw2023}, although physical mechanisms driving this increase likely differ.

\section*{Scaling law analysis}
Our extensive study with CR (see Fig.~\ref{fig:2}) challenges established results from microplastics and bacteria at several levels. First, we observe that a fixed and short bubble rising distance already leads to enrichment factors $E$ noticeably above unity, which cannot be explained by current surface-driven enrichment models. Moreover, $E$ further increases with $H$. Second, as we systematically varied the bulk concentration, an overlooked parameter, we reveal novel trends. Besides the regime of constant enrichment above unity at low bulk concentration, we uncover at higher concentration a new regime where the enrichment factor saturates, as it decreases with the bulk concentration in particles.

\subsection*{Constant enrichment factor}
\begin{figure}[t!]
\centering
\includegraphics[]{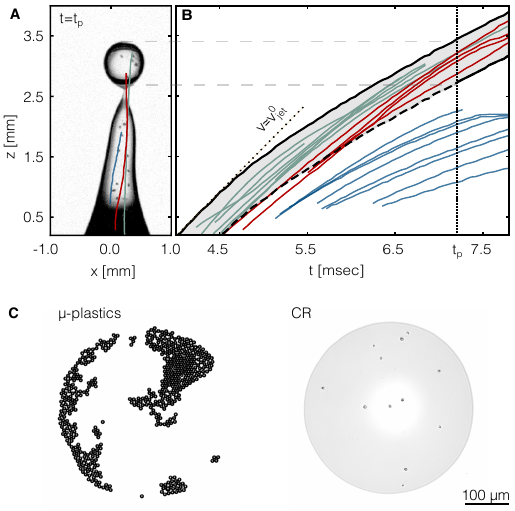}
\caption{Effect of particles' properties on the enrichment.
\textbf{A-B:} Tracking of OO cells trajectories inside a jet. 
\textbf{A:} Picture of the jet at pinching time $t_p$, when the first droplet detaches. Color lines highlight three cell trajectories until $t_p$. $z=0$ corresponds to the water free surface. 
\textbf{B:} Vertical position of cells as a function of time. The z-axis is shared with graph A. We choose $t=0$ at bubble burst. The upper solid black curve tracks the jet/droplet tip. The height of the gray area corresponds to the drop diameter. Green trajectories are always inside the grey area, red ones enter during the rise and blue ones stay outside the gray zone. The dotted line highlights the jet velocity at early times. 
\textbf{C:} Images of first drops with microplastics (left) and CR (right). For $C_{bulk} \approx 3.10^6$~cells/mL, $r_{d} = 0.3$~mm and $r_p = 5$~µm, microplastics' enrichment factor is two orders of magnitude larger than CR.
}
\label{fig:3}
\end{figure}

\paragraph*{Cells dynamics in the jet.}
To gain insights into a universal mechanism of enrichment that does not rely on surface effects, we capture high-speed image sequences of a jet. As CR are tiny algae (radius $r_p \approx 4$~µm, see SI), we follow the bursting of a millimetric bubble in an OO culture due to their larger size (see snapshot in Fig.~\ref{fig:3}A and Supplementary Movie 2). 
In Fig.~\ref{fig:3}B, we plot the vertical position of detectable cells in the jet of Fig.~\ref{fig:3}A as a function of time. The origin of time ($t=0$) is taken at bubble burst; the vertical line at $t_p$ shows the pinch-off time, when the drop separates from the jet (snapshot in Fig.~\ref{fig:3}A). The upper thick black line tracks the jet/drop tip, while the lower one follows the same dynamics, translated by $2r_d$, where $r_d$ is the first drop radius. The gray-shaded area between those two lines therefore highlights the portion of the jet which ends up in the first drop at $t_p$.

We categorize the trajectories into three distinct types, each plotted with a different color. Blue curves show cells which are not captured in the droplet. Green trajectories follow OO located within the gray-shaded region during the whole jetting process. Finally, red trajectories are the most intriguing: they track cells which initially lie outside the gray-shaded region but enter it before droplet pinch-off. Hence these extra cells increase the total number of cells in the first drop.

\paragraph*{Scaling model.}
Motivated by the red trajectories which highlights cells entering the droplet during the jetting process, we develop a toy model focusing on cells dynamics. 
We consider the jet tip (gray-shaded area in Fig.~\ref{fig:3}B) as a reservoir, fed with a jet of particles with cross-section $S_{neck} \approx \pi r_d^2$ until $t_p$. We assume that particles rise at the initial velocity of the jet $v_{jet}^0$, highlighted by the black dotted line in Fig.~\ref{fig:3}B. The particle flow rate $Q_p$ through the cross-section is then given by $Q_p \approx C_{bulk} v_{jet}^0 S_{neck}$. Thus, at $t=t_p$, the number of particles that successfully enter the reservoir is $N = Q_p t_p$ and the concentration in particles in the first drop, of volume $V_{drop}$, scales as $C_{drop} = N / V_{drop} \approx  C_{bulk} v_{jet}^0 S_{neck} t_p/V_{drop}$. 
Finally, the associated enrichment factor $E_c$ is given by:
\begin{equation}\label{eq:E}
	E_c \equiv \frac{C_{drop}}{C_{bulk}}\approx \frac{3 v_{jet}^0 t_p}{4 r_d}.
\end{equation}

To make further progress, we assume that water inertia, surface tension and viscous forces are of the same order of magnitude at the moment of cavity reversal. We also consider that the initial surface energy of the bubble supplies both the viscous dissipation in the capillary waves and the kinetic energy of jet formation. Then, we use the model of Gañ\'an-Calvo (2017) \cite{Ganan-Calvo2017} and also Lai et al. (2018) \cite{Lai2018} (see SI), who express $v_{jet}^0$ that reflects the initial flow (and therefore, cell speed), $r_d$ that relates to the drop formation and $t_p$ to the jet inertia. This leads to the extremely simple scaling for enrichment: 
\begin{equation}\label{eq:E*}
  E_c\approx k_e \left(\frac{l_{\mu}}{r_d}\right)^{1/5} \propto r_d^{-1/5} ,
\end{equation}
with the parameter $k_e\simeq36$ and the viscous length $l_\mu\equiv\mu^2/\rho\gamma$, where $\mu$, $\rho$ and $\gamma$ are the viscosity, density and surface tension of water. Therefore, $E_c$ remarkably only weakly depends on the drop radius $r_d$ and no other parameter. For droplet radii typically between 100 to 400~µm in our experiments, Eq.~\ref{eq:E*} predicts that $E_c$ is nearly constant, as we report. We estimate $E_c \approx 5$, without any adjustable parameter. Considering the scattering of our data Fig.~\ref{fig:2}, this result is in excellent agreement with our experiments. 

Our scaling analysis leads to the noteworthy simple predictive law of Eq.~\ref{eq:E*} that matches our regime of constant enrichment above unity. Besides, the dependency with the drop radius $r_d$ to the power $-1/5$ suggests a strong increase of the enrichment factor for very small drops. Experiments with microplastics and bacteria \cite{Dubitsky2023, Bezdek1972, Blanchard1970} highlighted this phenomenon, yet without a convincing explanation, which we now provide.

\subsection*{Saturation of the enrichment}
Our experiments at high concentration uncover a second regime: the decrease of the enrichment factor with $C_{bulk}$ (see Fig.~\ref{fig:2}B). We also evaluate the impact of cell size and shape on the enrichment, exploring various microalgae to vary the size over an order of magnitude and widen the ecological pertinence of our lab study. Whereas the cell body of CR is nearly spherical with a radius about 4~µm, OO is flat and oblong with a radius of 18~µm. We present complementary experiments (see SI) with \textit{Lingulaulax polyhedra} (LP): its size is similar to OO but its nearly spherical shape strongly differs. We also work with \textit{Proteomonas} sp. (PR), a fast swimming microalga with cell radius about 2~µm and a very oblong shape. These additional experiments emphasize how enrichment strongly depends on cell size, as the decrease of $E$ starts at lower $C_{bulk}$ for bigger cells. 

Remarkably, this decrease in enrichment does not seem to affect microplastics (see SI), which we demonstrate in Fig.~\ref{fig:3}C. We performed two bubble bursting experiments to compare microplastics and microalgae. We used hydrophilic microplastics with a radius of 5~µm, close to CR size. The release height ($H = 1.5$~cm), drop size ($r_d = 0.3$~mm) and bulk concentration ($C_{bulk} \approx 2.10^6$~cells/mL$^{-1}$, in the saturated regime) are identical in both cases. CR are dramatically less entrained than microplastics: enrichment factors are $E=3.4$ for microplastics and only $E=0.04$ for microalgae, two orders of magnitude lower. 

\begin{figure*}[t!]
\centering
\includegraphics{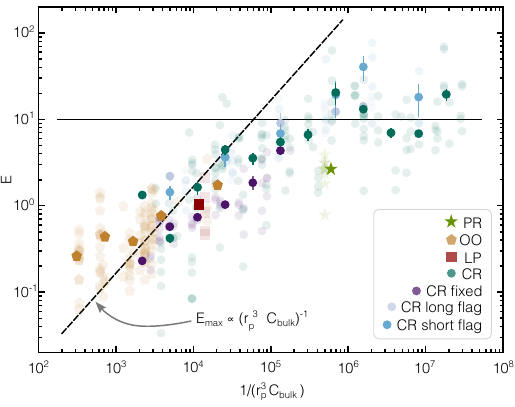}
\caption{Enrichment data from Fig. \ref{fig:2}B as a function of $1/(r_p^3C_{bulk})$. We added OO (orange pentagons), LP (red squares) and PR (green stars) as we vary the size and shape of the algae. The black solid line marks $E = 10$. The dashed line with equation $E = 1.7 \times 10^{-4}/(r_p^3 C_{bulk})$ represents the enrichment for maximum packing inside a sphere. Full dots are mean values of individual experiments shown in transparency, error bars are standard errors. }
\label{fig:4}
\end{figure*}

To model this marked decrease in microalgal enrichment factor, we consider cells as rigid spheres with radius $r_p$. The maximum number of spheres that can fit into $V_{drop}$ is $N_{max}=\phi \left(r_d/r_p\right)^3$,
where $\phi$ is the particle volume fraction. Thus the maximal enrichment factor follows:
\begin{equation}
	E_{max}=\frac{N_{max}}{V_{drop}C_{bulk}}=\frac{3\phi}{4\pi}\frac{1}{r_p^3C_{bulk}},
	\label{eq:Emax}
\end{equation}
where $E_{max}$ is inversely proportional to the bulk concentration at fixed $r_{p}$, as hinted by our measurements. This model also points that enrichment saturation is enhanced for large particle volume $r_p^3$, that is, a strong dependence on cell size. We test Eq.~\ref{eq:Emax} in Fig.~\ref{fig:4}, were we plot the enrichment factor of CR as a function of $1/(r_p^3C_{bulk})$, together with OO, LP and PR, which vary microalgal size and shape. Our data confirm the two distinct regimes: at low $1/(r_p^3C_{bulk})$, $E_{max}$ increases quite linearly as expected in the saturated regime and demonstrates that we captured the size dependency, while for larger value we reach the constant regime with $E \simeq 10$.

Finally, as we adjust Eq.~\ref{eq:Emax} to our data in Fig.~\ref{fig:4} with a coefficient $1.7 \times 10^{-4}$, we get an estimate of the effective particle radius $r^*$. Considering random packing with $\phi \approx 0.64$, we evaluate $r^* \simeq 10 r_p$. Therefore, microalgal apparent radius is considerably larger than their own body size. This does not apply to microplastics, as we highlight in Fig.~\ref{fig:3}C that microplastics remain in the constant regime when CR is already in the saturated one. To evaluate $r^*$ in microplastics, we aimed for their maximal packing regime by further increasing bulk concentration. We failed due to their strong aggregation that led to sedimentation. We conclude that the effective radius of microplastics is close to their native $r_p$, while microalgae naturally repel, increasing $r^*$ their effective radii. This large difference between both apparent radii suggests that particles are selectively sorted during bursting, depending on their properties and the jet hydrodynamics, a process that remains to be understood.

\section*{Discussion and perspectives}

Studies have shown that algal cells and cell fragments can be present in sea spray \cite{Veron2015, Alsante2021}, which may contribute to airborne toxicity in the case of certain harmful algal blooms like those of \textit{Ostreopsis} spp. Despite this potential relevance to human health, little is known about the aerosolization process for microalga and their potential enrichment in sea spray. 
Our extensive set of experiments uncovers two regimes for the enrichment factor as a function of the concentration, an overlooked but ecologically pertinent parameter. Field measurements during harmful algal blooms validate the range of concentrations we explored \cite{Ternon2025, Murray2026, Berdalet2022, Chomerat2022}. At low concentration, in the constant regime, our results agree at first sight with the literature on bacteria and microplastics as $E$ does not depend on $C_{bulk}$ but increases with $H$, the bubble rising distance in the water column \cite{Blanchard1981, Shaw2023, Ji2022}. The literature justifies this later dependency by surface accumulation of particles captured at the interface. 
Yet, this explanation faces two major limitations. First, over short rising distances, the bubble surface does not scavenge many particles, while enrichment factors for microplastics or microalgae are indubitably larger than unity. Second, entrapment mechanisms for large particles such as OO or LP are still a mystery. 

Here, we experimentally demonstrate the mechanisms behind this unsolved conundrum with our minute high-speed observation of the jet (Fig.~\ref{fig:3}A-B and Supplementary Movie 2). We demonstrate that surface accumulation is not required to observe $E > 1$. Indeed, cells within the jet rise faster than the jet tip, which decelerates due to gravity. As a consequence, microalgae accumulate at the top and the first drop eventually pinches-off, highly loaded with particles. We derive a scaling analysis based on this particle flow in Eq.~\ref{eq:E*}. Our simple model predicts that enrichment only weakly varies with the drop radius for the range of bubble sizes we explored, $E \propto r_d^{-1/5}$. This law therefore catches two essential behaviours: it predicts both extreme enrichment factors at very small drop radii \cite{Bezdek1972, Dubitsky2023, Blanchard1970} and a nearly constant enrichment for slightly larger droplets, in excellent agreement with our experiments and published results on microplastics \cite{Shaw2023}. Additionally, this hydrodynamic explanation of enrichment factors above unity justifies the entrapment of large particles, giving new clues in a long-running debate in the literature \cite{Blanchard1989, Ji2022, Dubitsky2023}. 

We took advantage of the large size of OO to qualitatively observe it inside a jet to gain a better understanding of jet dynamics and the enrichment of large particles. Even if we did not reach the constant enrichment regime for OO, these experiments highlighted the complex flows inside the jet, leading to cells rising faster than the jet tip itself. Indeed, flows driving the jet rise have a rich dynamics that leads to a larger velocity at the jet center \cite{Deike2018, Hoepffner2013}. This velocity gradient across the jet then locally shears microalgae. As CR is non-spherical, owing to its two forefront flagella, it is efficiently reoriented by shear and flow-focused at the center of the jet \cite{Kessler1986}. Thus, as the core of the jet is more concentrated and rises faster, the reservoir-filling model describes very well the constant enrichment regime of CR. Consequently, our simple scaling offers a satisfactory description of the enrichment factor for microalgae ranging from 5 to 20~µm in radius and also extends to microplastics ranging from 5 to 140~µm \cite{Dubitsky2023, Shaw2023}.

Furthermore, we predict a saturation regime at high cell concentration in the water column. Under the assumption of maximal packing, $E$ inversely scales with both $C_{bulk}$ and the particle volume. Our experiments at high concentrations show this decrease in enrichment factor, which was not reported for either microplastics or bacteria due to a lack of exploration of this parameter. We attempted to complement results for microplastics, but failed as they aggregate and sediment before reaching the maximum packing condition. Adjusting Eq.~\ref{eq:Emax} to our data in Fig.~\ref{fig:4}, we evaluate that CR and OO are well described by an effective radius $r^*$ about 10 times larger than their body size. 
As these microalgae are nearly immobile during the very short timescale of bubble bursting, hydrodynamic repulsion is unlikely. As living cells carry a negative surface charge, electrostatic repulsion may also be assumed to participate in cell-cell interactions in this context \cite{Li2022}. However, electrostatic interactions are not sufficient to explain the increase of an order of magnitude in $r^*$, leaving the physical origin of this apparent radius an open question.  

To challenge the universality of our model built with CR and OO, two microalgae of differing size and shape, we confronted it to extra microalgae. 
LP is more spherical than OO but of similar size, and the corresponding experiments fall well within the validity of our model. We note however that enrichment factors for PR (green stars in Fig.~\ref{fig:4}) slightly deviate from our predictive laws: it is the most oblong, tiniest and much fastest microalga of our set, and these characteristics may prevent it to meet all our model hypotheses.

Finally, the mechanisms we uncover behind enrichment of microalgae in jet drops stimulate curiosity about the potential enrichment of spume and film drops, where other physical phenomena may be at play. This would provide a comprehensive picture of the global sea spray enrichment dynamics, where hydrodynamics plays a major role. The mechanisms that we evidence here demonstrate how very large particles can be entrapped in sea spray, driving the transfer of ocean-extracted biological matter to the atmosphere. Our predictive laws also advance the understanding of how toxin-bearing cells and cellular fragments can become airborne, revealing the need to further investigate the role of particulate matter in airborne toxicity during harmful algal blooms.

\end{refsegment}
\printbibliography[segment=1, check=onlynew]

\section*{Methods}
\begin{refsegment}

\subsection*{Single bubble bursting setup} 
We followed the dynamics of a single bubble burst at the time. For this purpose, a submerged needle blew air in a small box filled with an algal culture at concentration $C_{bulk}$. We built home-made cubic boxes of 2 or 3~cm side with PMMA, with a needle going through the bottom plate. We adjusted the height $H$ between [1.5, 10]~cm. For higher boxes with $H$ up to 25~cm, we widened the box up to 5~cm sides to avoid wall effects, which we controlled with the dynamics of the bubble and jet (see SI). These heights are comparable to studies with microplastics and bacteria \cite{Blanchard1989, Ji2022, Shaw2023}. We used needles with different internal diameters, from $0.08$ to $1.37$~mm, to vary the bubble size from $0.25$ to $2.5$~mm, as it controls the ejected droplet radius \cite{Ganan-Calvo2017, Blanchard1989, Dubitsky2023} (Techcon TE needles, internal diameter $\in$ [0.08, 1.37]~mm). For smaller bubbles, we heated and stretched glass capillary tubes (Hirschmann ringcaps 9600110), fabricating our own thin glass needles. 

A syringe pump (WPI AL-1000) pushed air through the needle, forming a bubble that subsequently detached and rose to the liquid surface. Once at the surface, we measured the bubble size with a camera (Nikon D800). For large bubbles with an ellipsoidal shape, we evaluated the equivalent bubble radius $R_b=(a^2b)^{1/3}$ from side pictures, where $a$ and $b$ are the major and minor axis of the ellipsoid, respectively. A side high speed camera (Phantom v2511 or Phantom VEO 640L, frame rate from 2500 to 40000 fps) filmed single bubble burst and jet production as in Fig.~\ref{fig:1}C-D, giving access to first drop velocity and radius. We discarded jets with an angle to the vertical above 3°. To image without phototactic effects, we illuminated with a red LED (Advanced Illumination CX0404-625IC) \cite{Foster1980}. We systematically mixed the algal cultures before each bubble burst to ensure a uniform distribution of cells throughout the box. 

We checked that neither our small boxes nor the presence of microalgae impaired the drop size and dynamics (see SI). We collected the first and fastest drop on a glass slide, whose height was adjusted to capture solely this drop. We observed this single droplet under the microscope (inverted microscopes Leica DMi8 or Zeiss Axiovert 40 CFL) to count the number of cells, illustrated Fig.~\ref{fig:1}E.

We measured the equilibrium surface tension of all cultures and solutions with a Du Noüy ring tensiometer (KRÜSS K6), and verified it is close to water ($\gamma\approx$~70~mN/m). The bulk viscosity (measured with an Anton Paar Lovis 2000 M/ME Rolling Ball Viscosimeter) of each solution was also close to water. 

To explore a wide range of concentrations, we either filtered or centrifuged cultures, or we diluted algae in their own culture medium until no cells are caught in the first drop across repeated measurements. Microplastics were polystyrene beads with density $\rho_p=$ 1050~kg/m$^3$ and radius $r_p=$ 5~µm (microbeads dynoseeds TS-10).

\subsection*{Microalgae} 
\paragraph*{Chlamydomonas reinhardtii}
Most of our experiments involved the non-toxic unicellular microalga \textit{Chlamydomonas reinhardtii}, a well-characterized freshwater model microorganism~\cite{Goldstein2015, Jeanneret2016, Salome2019, Harris2009}. Wild-type CR has a body diameter of approximately 8~µm (see histogram and picture in the SI) and two flagella about 10~µm in length. This algae typically swims at 100~µm/s. 
We grew three \textit{C. reinhardtii} strains: CC-125 WT, CC-2289 lf3-2 mt- (long flagella) and CC-3663 shf1 mt+ (short flagella) from the Chlamydomonas Center, Minnesota, axenically in Tris-acetate-phosphate (TAP) at 22°C inside an incubator (Memmert IPP260ECOPLUS) programmed with a diurnal cycle of 14:10 hours of light (6.4 W/m$^2$) and darkness. We grew 50 mL cultures in 250 mL Erlenmeyers, under constant agitation at 120 rpm (Heidolph Rotamax 120 orbital shaker). We sampled cells during their exponential growth phase (from 3 to 8 days after inoculation to maximize algal fitness). To fix cells, we added 1\% of Lugol (Sigma-Aldrich 32922) to cell cultures. We evaluated cell concentration with an improved Neubauer counting chamber.

\paragraph*{Ostreopsis cf. ovata}
We cultured two strains of \textit{O}. cf. \textit{ovata}: V-OST-4 which was isolated from a macroalga sampled in Ocata (Spain, NW Mediterranean) in September 2017 and V-OST-5 which was isolated from a macroalga sampled in Deux-Jumeaux (Bay of Biscaye) in September 2023. They were maintained at 22°C in an incubator (Memmert 260 ECOPLUS) programmed with a diurnal cycle of 14:10 hours of light (60 µmol photons m$^{-2}$s$^{-1}$) in L1/2 medium. For the purpose of the study, we cultured them in 800 mL Fernbach flasks to offer a large settling surface for the benthic cells. This microalga is flat and oblong, with a cell body diameter about 35~µm (see histogram and picture in the SI). Cellular concentration was determined using a 1 mL Sedgwick Rafter Counting Cell.

\paragraph*{Lingulaulax polyedra}
\textit{L. polyedra} (LP) is a biflagellated dinoflagellate also associated with reports of respiratory irritations in California \cite{Ternon2025}. It has a nearly spherical shape with a body diameter close to 40~µm (see histogram and picture in the SI). We cultured strain V-LIN-1 (RCC11581) in the same conditions as \textit{O.} cf. \textit{ovata} strains although we use L1 medium and 5-L-erlenmeyer. Cellular concentration was also determined using a 1 mL Sedgwick Rafter Counting Cell.

\paragraph*{Proteomonas sp.}
\textit{Proteomonas} sp. (PR) is a small marine cryptophyte, with a body diameter of approximately 5~µm (see histogram and picture in the SI). This elongated biflagellated organism is not currently associated to respiratory effects, and it is a non-harmful algal bloom species. Yet, the tiny rugby ball shape offers an extension of the physical parameters of our study. The \textit{Proteomonas} strain is not monoclonal and was recently isolated from Roquebrune Cap-Martin using flow cytometry (07/07/2025). The culture was grown at 19°C in an incubator (Sanyo MLR-351) programmed with a diurnal cycle of 14:10 hours of light (60 µmol photons m$^{-2}$ s$^{-1}$) in F/2 medium. They were cultivated in 5-L-erlenmeyer in which we observed the formation of a thick biofilm. Cellular concentration was determined using an optical particle counter (PAMAS OLS).

\subsection*{Field experimental setup}
Sampling for airborne \textit{Ostreopsis} cells was performed at high tide, during an established bloom of \textit{Ostreopsis} cf. \textit{ovata}, on July 9th and 10th, 2025 at the Erromardie beach and on July 16th, 2025 at the Mayarco beach, in Saint Jean de Luz, France (43.4072, -1.64143 and 43.4186, -1.62518, respectively). A Coriolis µ biological air sampler (Bertin Technologies) was installed 20 meters from the shoreline to collect air at a flow rate of 300 L/min for 30 min, giving a total sampled air volume of 9000 L per sample. The collection cones were filled with 15 mL of 0.2 µm filtered 36 PSU seawater. After collection, the samples were immediately fixed with acidic lugol’s solution (1\%, v/v) and transferred into glass vials before storage at 4°C in the dark until processing. In the lab, sample volumes were adjusted to 50 mL with sterile seawater, after which samples were transferred to a sedimentation column and left undisturbed for 24h before being counted according to Utermöhl \cite{Utermohl1958}. Note that SW sampling was performed in the morning at low tide (see SI), therefore cell concentrations in SW may not be representative of the actual cell content at the time of the air sampling.

\section*{Acknowledgement}
This project received financial support from the CNRS through the MITI interdisciplinary programs (EFT, HM).
This work was supported by the project OSTREOBILA led by CAPB and funded by FEDER (ERDF) and INTERREG POCTEFA programmes (ET).
We thank the GDR OMER which favored this collaboration. 
We thank Pier Giuseppe Ledda for discussion, Matthieu Roché who gave us microplastics. We thank Stéphane Popinet, Axel Huerre, Francesco Picella and Mary Kane for proofreading the manuscript.

\section*{Author contributions}
EFT, JP and HM conceived and designed the study. EFT built the setup, performed all laboratory experiments and their physical analysis, JP and HM supervised the work. EFT, JP and HM interpreted the results. MDR and ET did the field work and related analysis. PC and ET provided OO, LP and PR cultures and contributed to the bubble bursting experiments. All authors wrote the paper.

\section*{Competing interests}
The authors declare no competing interests.

\section*{Online content}
Supplementary information and movies are available for the review process. SI datasets will be made available online after acceptance.

\end{refsegment}
\printbibliography[segment=2, check=onlynew]

\newpage

\section*{Supplementary Information}
\begin{refsegment}

\subsection*{Content}
\begin{itemize}
    \item Movie 1: High speed movie of cavity collapse after bubble bursting. 
    \item Movie 2: Jet resulting from bubble bursting in a culture of OO. Cells are visible as black dots. 
    \item SI data set (after acceptance).
    \item Supplementary document, including additional information, figures and tables: see below. 
\end{itemize}

\subsection*{Ecological implication}

We performed field measurements in 2025 during an \textit{Ostreopsis} cf. \textit{ovata} (OO) bloom occurring along the French Atlantic coast, on July 9th and 10th, 2025 at the Erromardie beach and on July 16th, 2025 at the Mayarco beach, in Saint Jean de Luz, France (43.4072, -1.64143 and 43.4186, -1.62518,  respectively). Specifically, planktonic OO cell concentration at low tide in sea water (SW) scaled about $30$~cells/mL with peaks up to $600$~cells/mL (2025, Basque country; see Table~\ref{tab:ecologicaldata}), a value triggering respiratory syndromes. The full description of the experimental setup is given in the Methods. We report three days of measurements in Table~\ref{tab:ecologicaldata}: OO cells and ruptured OO cells are systematically captured in the collector. As we did not have access to bubble sizes, we cannot evaluate natural enrichment factors. 

In the context of harmful algal blooms (HABs) inducing respiratory irritations, cell concentration in the water column may reach up to $10^3$ to $10^5$ cells/mL as seen for recent blooms of \textit{Lingulaulax polyedra} (LP) (2020, California \cite{Ternon2025}), \textit{Karenia cristata} (2025, Australia \cite{Murray2026}), OO (various years and location \cite{Berdalet2022, Chomerat2022}). Note that cell concentrations of OO are larger in the sea surface microlayer than in the water column, stressing how cell coverage is intrinsically heterogeneous in the sea and may form high-cell density patches at the sea surface \cite{Berdalet2022, Ternon2024}. Therefore, the bulk concentrations we explore in this study are ecologically relevant.

\subsection*{Experimental setup}
\begin{figure}
\centering
\includegraphics[width=\textwidth]{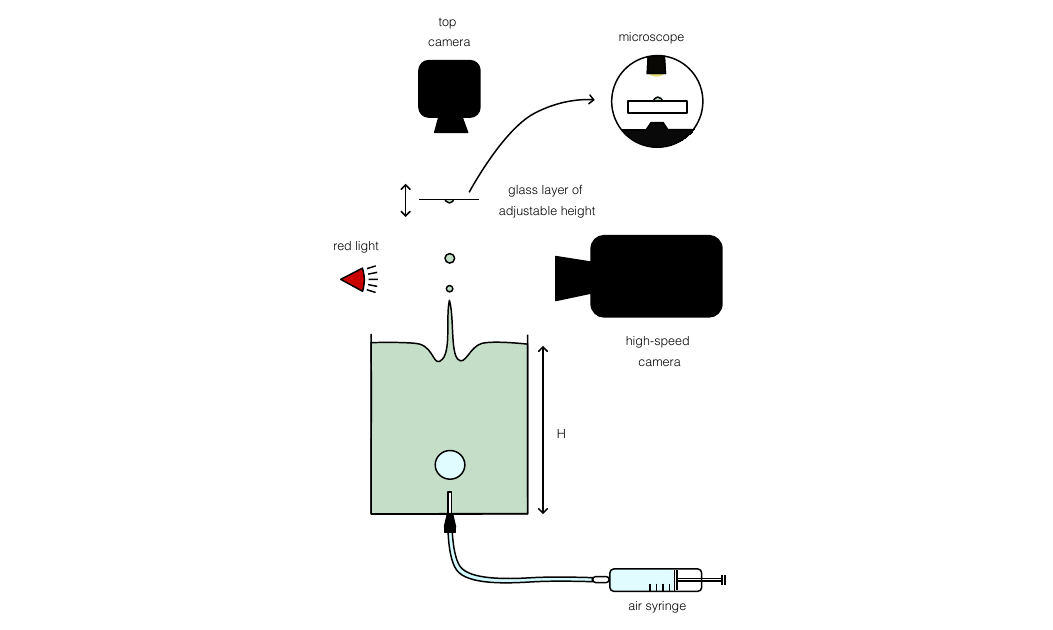}
\caption{Schematic of the experimental setup.}
\label{fig:exp_setup}
\end{figure}

We depict the experimental setup given in the Methods, in Fig \ref{fig:exp_setup}.

\subsection*{Surface tension of cultures}
\begin{figure}
\centering
\includegraphics[width=\textwidth]{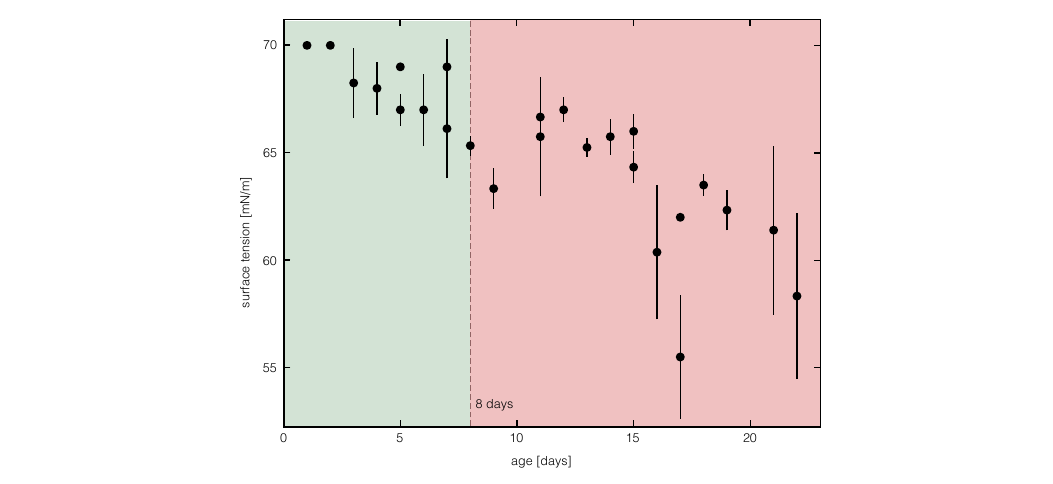}
\caption{Surface tension of CR cultures as a function of culture age. Black dots are mean value of repeated measures (typically 5 essays) on a single solution. Error bars corresponds to standard deviation. For our purposes, we used cultures less than 8 days after the inoculation.}
\label{fig:surface_tension}
\end{figure}

Since higher concentration implies more biomass, we checked if culture ageing had an impact on surface tension. Fig. \ref{fig:surface_tension} shows liquid surface tension of \textit{Chlamydomonas reinhardtii} (CR) cultures as a function of culture age: we performed repeated measurements on several cultures, reported as mean values (black dots) and standard deviations (error bars). Although surface tension decreased with culture age, the value did not deviate significantly from that of water for cultures of less than 8 days.
Note that we obtained low algal concentrations by adding culture medium, which does not reduce the surface tension of the suspensions.

\subsection*{Drop size and dynamic}
We compared our experiments to the literature to validate the use of established scaling for simple liquids. We systematically measured geometrical parameters and dynamical characteristics. In this process, we verified that the presence of microalgae did not modify the physico-chemical properties of the liquid, and that the size of the box, chosen small to minimize cultures' volumes, did not impair the drop size and dynamics. 

\begin{figure}
\centering
\includegraphics[width=\textwidth]{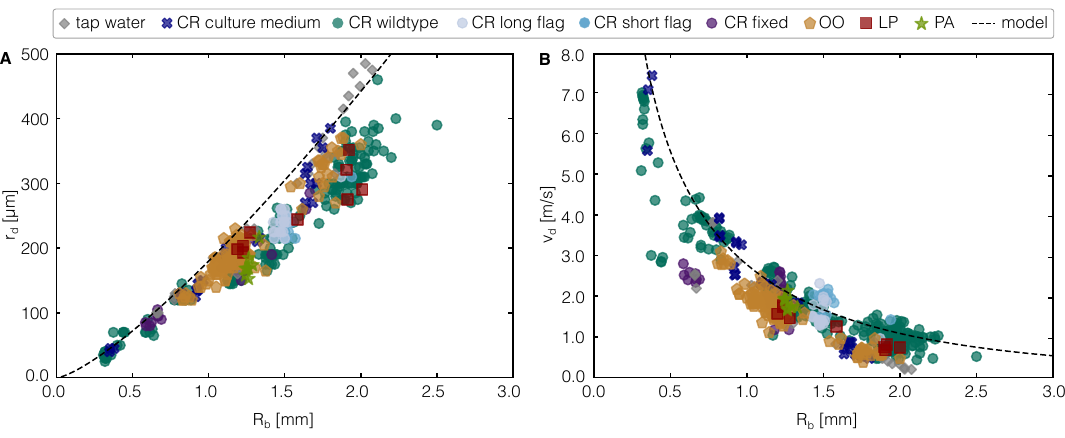}
\caption{Radius $r_d$ (\textbf{A}) and velocity $v_d$ (\textbf{B}) of the first jet drop as a function of the bubble radius $R_b$. Experimental points are compared with SI equation~\ref{eq:rd} and \ref{eq:vd}.} 
\label{fig:dynamic}
\end{figure}

Fig. \ref{fig:dynamic} shows our measurements of drop radius $r_d$ and velocity $v_d$ as a function of the bubble radius $R_b$. Drop velocity was measured by tracking the drop trajectory immediately after pinch off. We performed experiments with tap water, CR culture medium TAP and with cultures of microalgae.

We compare experimental results with scaling laws for the dynamic of bubble bursting of simple liquids. Let $\varphi \equiv \sqrt{\text{La}} \left(\sqrt{\text{La}/\text{La}_\star}-1\right)$ be a dimensionless parameter, where La~$=\rho\gamma R_{b}/\mu^2$ is the Laplace number (with $\rho$, $\gamma$ and $\mu$ the density, surface tension and viscosity of the liquid), and La$_\star$~=~500 is the critical Laplace value below which no droplet detach. Then, Gañ\'an-Calvo (2017) \cite{Ganan-Calvo2017} and also Lai et al. (2018) \cite{Lai2018} expressed the vertical initial jet velocity $v_{jet}^0$, drop radius $r_d$ and pinching time from the jet formation $t_p-t_0$ as follow:
\begin{align}
	v_{jet}^0/V_{\mu} = k_v \varphi^{-3/4} \label{eq:vj}\\ 
    r_d/l_{\mu} = k_d \varphi^{5/4} \label{eq:rd}\\ 
    (t_p-t_0)/(l_\mu/V_\mu)=k_t \varphi^{7/4} \label{eq:tp}, 
\end{align}
where $V_\mu\equiv\gamma/\mu$, $l_\mu\equiv\mu^2/\rho\gamma$ and $k_v\simeq16$, $k_d\simeq0.6$ and $k_t\simeq2$ are fitting parameters. We use these equations to derive the final expression of the enrichment factor in Eq. \ref{eq:E*} of the main text.

Dashed curves in Fig. \ref{fig:dynamic} show the scaling derived for water ($\gamma=70$ mN/m, $\rho=10^3$ kg/m$^3$ and $\mu=10^{-3}$ Pa.s). In Fig. \ref{fig:dynamic}A we trace the model established by Gañ\'an-Calvo (2017) \cite{Ganan-Calvo2017}, as reported in Eq. \ref{eq:rd}. In Fig. \ref{fig:dynamic}B we trace the expression of first drop velocity $ v_{jet}^0$ as improved by Deike et al. (2018) \cite{Deike2018}, taking into account the effects of gravity:
\begin{equation}
    v_{jet}^0=k_v\frac{\gamma}{\mu} \left(\left(1+\alpha \text{Bo}\right)\text{La}\right)^{-3/4}\left(\frac{1}{\sqrt{\text{La}_\star}}-\frac{1}{\sqrt{\text{La}}}\right)^{-3/4}
      \label{eq:vd}
\end{equation}
where Bo~$=\rho g R_b^2/\gamma$ is the Bond number, and $k_v=19$ and $\alpha=2.2$ are fitting parameters. Note that this prediction refers to $v_{jet}^0$, the velocity of the jet tip at cavity reversal, which slightly differs from the velocity of the jet tip (i.e., of the droplet) that we measured immediately after the first drop detaches. 

Since our experimental data align with theoretical predictions, we conclude that neither the microalgae nor the size of the boxes significantly affects the relationship between bubble size and the size and velocity of the first drop.

\subsection*{Algal species and CR mutants}
\begin{figure}
\centering
\includegraphics[width=\textwidth]{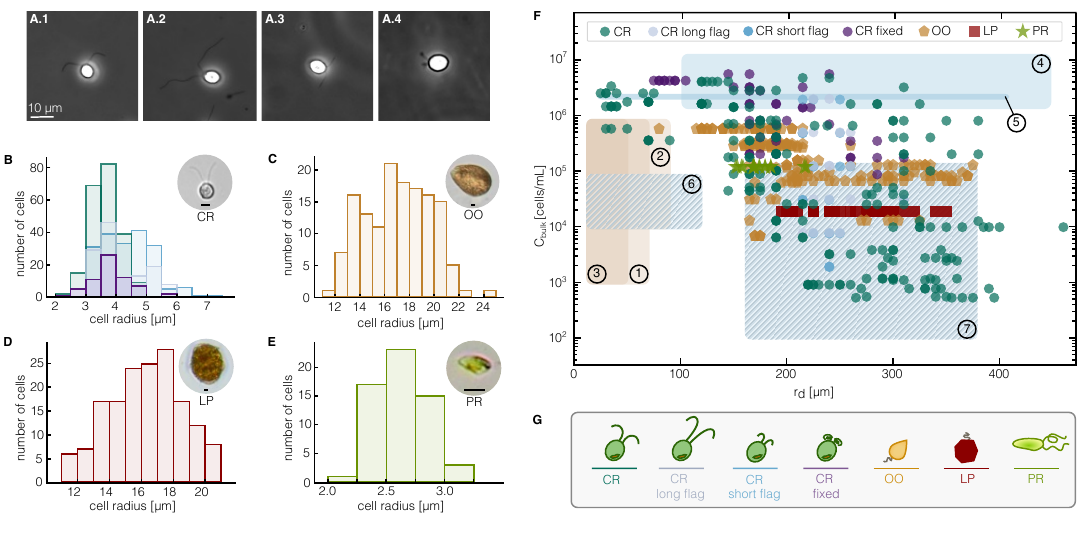}
\caption{Microalgae characteristics. 
\textbf{A}: Phase-contrast-microscopy images of CR. Cells are sourced from the Chlamydomonas Center, Minnesota. The scale bar shows 10 µm and stands for all the pictures. \textbf{A.1:} CC-125 wild-type. \textbf{A.2:} CC-2289 lf3-2 mt- (long flagella). \textbf{A.3:} CC-3663 shf1 mt+ (short flagella). \textbf{A.4:} CC-125 wild-type fixed with 1\% lugol solution.
\textbf{B-E:} Histograms of cell sizes on each microalgae specie.  \textbf{B}, \textbf{C}, \textbf{D} and \textbf{E}  correspond respectively to CR, OO, LP and PR, illustrated in the related pictures. All scale bars show 5~µm. 
\textbf{F:} Phase space of the bulk concentration $C_{bulk}$ as a function of the drop radius $r_{d}$. We compare literature results (colored squares, brown for bacteria, blue for microplastics) to our experiments with $H \in [1.5,\,2]$ cm. Hashed areas show studies with the closest parameters to our work. Number labels refer to Table \ref{tab:litterature}. 
\textbf{G:} Schematic catalog of the microalgae of this study: wild-type CR, long and short flagella mutants CR, fixed wild-type CR, OO, LP and PR. Not to scale. }
\label{fig:species}
\end{figure}

Most of our experiments involve the non-toxic unicellular microalga \textit{Chlamydomonas reinhardtii} (CR), a well-characterized freshwater model microorganism~\cite{Goldstein2015, Jeanneret2016, Salome2019}. CR offers access to a wide range of mutants, a well-documented life cycle \cite{Harris2009}, and a comprehensive understanding of its motility~\cite{Polin2009, Leptos2023} as well as its hydrodynamic interactions with the surrounding fluid and rigid boundaries~\cite{Kantsler2013, Contino2015, Lushi2017, Mirzakhanloo2018}. Wild-type CR has a body diameter of approximately 8~µm (see picture and histogram in Fig.~\ref{fig:species}A.1 and Fig.~\ref{fig:species}B) and two flagella about 10~µm in length.  

In addition, we used short- and long-flagella mutants and fixed wild-type CR cells. Fig.~\ref{fig:species}A shows phase-contrast microscopy images of cells, where we estimate the flagella length $l$: wild-type ($l \approx 12$~µm), long ($l \approx 20$~µm) and short-flagella ($l \approx 7$~µm) mutants and fixed ($l \approx 3$~µm) cells.

We complemented our investigations with three marine algal species of diverse body sizes and shapes, in order to test the ecological relevance of our physical model. We provide in Fig.~\ref{fig:species}C-E histograms and pictures of \textit{Ostreopsis} cf. \textit{ovata}, \textit{Lingulaulax polyedra} and \textit{Proteomonas} sp.

\subsection*{Parameters' space}
We present in Fig.~\ref{fig:species}F our phase space involving bulk concentrations $C_{bulk}$ and droplet radii $r_d$, which we compare to the existing literature on single bubble bursting in particulate water. We summarize this literature and all microalgae of this study in Table~\ref{tab:litterature} and Fig.~\ref{fig:species}G. To explore a wide range of concentrations, we either filter or centrifuge cultures, or we dilute algae in their own culture medium until no cells are caught in the first drop across repeated measurements. Hatched regions highlight the two studies in which particle sizes are close to those investigated in this work~\cite{Dubitsky2023, Shaw2023}.

\subsection*{Enrichment as a function of bulk concentration}
Fig. \ref{fig:Enrichment}A shows the enrichment as a function of bulk concentration for all algal species. The behavior of CR is already presented in the main text.

Experiments with PR (light green stars) have a concentration $C_{bulk} = 10^5$~cells/mL, within the transition zone between the two regimes highlighted with CR. Enrichment factors from our smallest specie are slightly below those of CR. Cell body sizes are in the same range, but their shape strongly differs and flagellar extent varies between the two species, which may play a role on a passive reorientation mechanism. 

LP and OO are of similar equivalent radius--both about 5 times larger than CR--although their shapes strongly differ. Yet, our few LP data (red squares) are consistent with the large campaign we performed with OO (orange pentagons). Enrichment factors are systematically 10 times smaller for theses species than for CR, which eventually highlights the clear impact of the cell body size on the final droplet enrichment. Note that these enrichment factors are of order 1 or smaller. Yet, OO's enrichment decreases as the bulk concentration increases, following the same trend as CR. We also expect a regime of constant enrichment at smaller concentration.

\subsection*{Enrichment as a function of drop radius}
Previous works on bacteria and microplastics \cite{Blanchard1989, Dubitsky2023} show a strong dependency of the enrichment factor $E=C_{drop}/C_{bulk}$ with the size of the droplet (i.e., of the bubble). As we vary $r_d$ from 50 to 500~µm in Fig.~\ref{fig:Enrichment}B, we do not observe any significant dependency of the enrichment for wild-type CR. Note that results are color-coded with the bulk concentration, stressing how data for $C_{bulk} > 10^4$~cells/mL strongly deviate because they enter the saturated regime. We added results for $N_{cells} = 0$ at the bottom of the logarithmic vertical axis. 
The line shows Eq.~2 from the main text, highlighting the weak dependency of $E$ in $r_d^{-1/5}$: the corresponding increase at small $r_d$ is predicted for smaller radii than what we explored, and the order of magnitude agrees well with our results.

\subsection*{Enrichment as a function of bubble rising distance}
Many studies in the literature \cite{Blanchard1981, Ji2022, Shaw2023} report from experiments that the enrichment factor increases with the distance $H$ traveled by the bubble through water: while rising, the surface scavenges bacteria and particles. This leads to $E\propto(H/R_b)^{\alpha}$ where $\alpha$ depends on assumptions on particle size and bubble surface properties: $\alpha=1$ for free-slip and 1/2 for no-slip conditions \cite{Dai2000}.

In order to explore enrichment dependency on bubble rising distance, we built boxes of different heights to change $H$ the travel distance through the liquid from 1.5 up to 25~cm.

We plot in Fig.~\ref{fig:Enrichment}C the enrichment $E$ as a function of $H$ for wild-type CR. We kept $C_{bulk} \leq 10^4$~cells/mL to avoid any effect of the concentration. The enrichment increases almost proportionally with the rising distance, that is, $\alpha \approx 1$.

\subsection*{Comparing microalgae and microplastics}
\begin{figure}
\centering
\includegraphics[width=\textwidth]{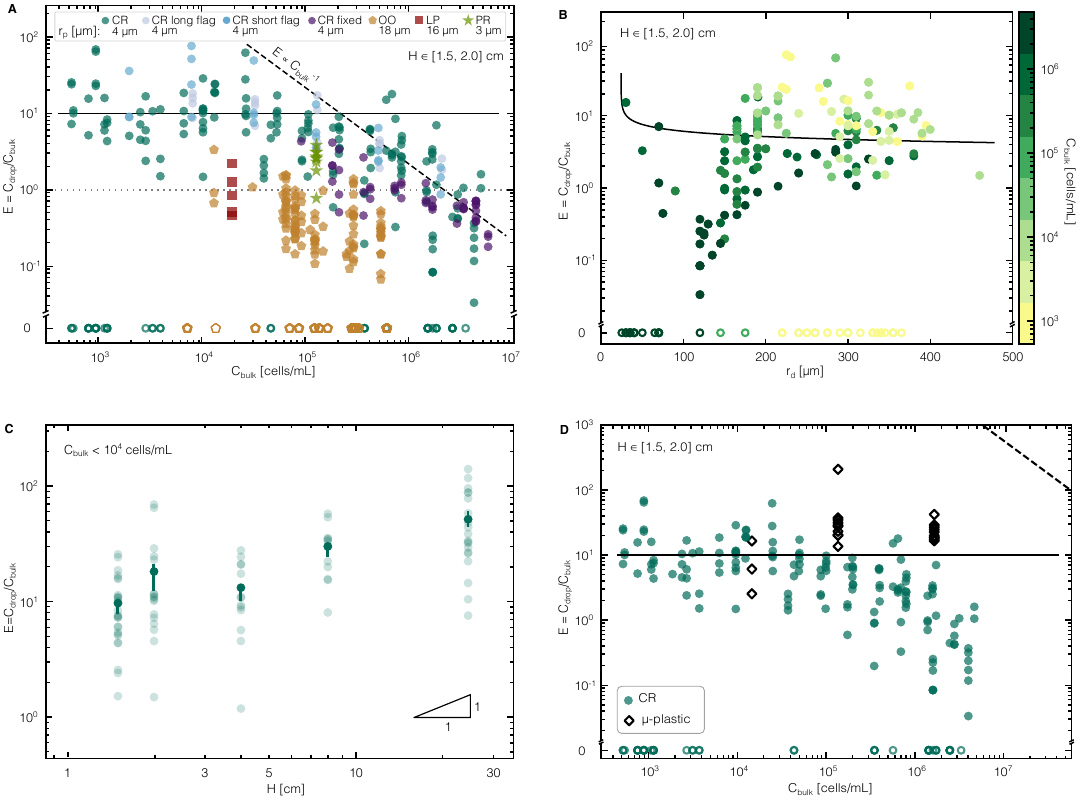}

\caption{Enrichment factor. 
    \textbf{A:} Enrichment $E$ in the first drop as a function of the bulk concentration $C_{bulk}$ for all microalgae species at a fixed bubble rising distance $H \in[1.5,\,2]\,$cm. The black solid line marks $E=10$, the dashed line represents $E\propto C_{bulk}^{-1}$, the grey dotted line marks $E=1$. Dots show CR results, dark green: wild-type CR, light gray: long flagella CR mutant, blue: short flagella CR mutant, purple: fixed wild-type cells. Orange pentagons: OO, red squares: LP, green stars: PR. Measurements with $E = 0$ are reported at the bottom of the split logaritmic vertical axis. 
    \textbf{B:} Enrichment as a function of drop radius for wild-type CR with $H \in [1.5, 2]$~cm. Color code stands for bulk concentration. The black line shows the Eq. 2 in the main text. We include experiments with $N_{cells}=0$ as empty circles by splitting the logarithmic vertical axis. 
    \textbf{C:} Enrichment $E$ for CR in the first drop as a function of the bubble rising distance $H$, for $C_{bulk}<10^4\,$cells/mL. Opaque dots are mean values, error bars are {standard errors}. All experimental points are reported in partial transparency. 
    \textbf{D:} Enrichment $E$ as a function of bulk concentration $C_{bulk}$ for CR wild-type (green dots) and microplastics of $r_p=5$~µm (black diamonds). The solid line shows $E=10$. The dashed line in the upper right corner indicates the maximum enrichment for rigid spheres of radius $r_p=5$~µm randomly ($\phi=0.64$) packed in a larger sphere of radius $r_d$. We include experiments with $N_{cells}=0$ as empty circles by splitting the logarithmic vertical axis. 
}

\label{fig:Enrichment}
\end{figure}

We demonstrate in the main text that the enrichment factor decreases at high bulk concentration in microalgal cultures. We attribute this trend to a maximum packing effect, which limits the number of cells that can be accommodated within the droplet. Nevertheless, if we calculate the maximum packing of small spheres (i.e, cells) inside a larger sphere (i.e, the droplet), we find that microalgae effectively occupy a volume much larger than their body volume, as if cells have an effective radius $r^*>r_p$. To our knowledge, this effect has not been reported for microplastics. Yet, the concentration was not explored systematically. Therefore, we performed a few experiments with polystyrene ($\rho_p=1050$ kg/m$^3$) spherical microplastics of $r_p=5$~µm, i.e., of size and density comparable to CR. Figure \ref{fig:Enrichment}D shows the enrichment as a function of the bulk concentration, and we compare microplastics (black diamonds) and wild-type CR (green dots). The black dashed line in the upper right corner shows the maximum enrichment factor for particles of $r_p=5$~µm randomly packed ($\phi=0.64$) in a sphere, $E_{max}=3\phi/(4\pi r_p^3C_{bulk})$. While $E$ starts decreasing from $10^5$ cells/mL in microalgal cultures, our experiments with microplastics demonstrate that $E$ stays constant and about 20 in our experiments, for $C_{bulk}$ up to $2.10^6$ particles/mL. We tried to go beyond this last concentration, but were not able to reach the maximum packing limit indicated by the dashed line: the sedimentation time of the microplastics becomes comparable to the duration of the experiment. Particles both significantly sank and formed clusters that accumulated at the water surface. Due to these visible artefacts, further experiments would not be comparable to our earlier results. 
Yet, we demonstrate that microplastics follow another aggregation dynamics, and that their effective radius is close to their intrinsic radius.

\end{refsegment}
\printbibliography[segment=3, check=onlynew]

\newpage

\begin{sidewaystable}[t!]
\centering

\begin{tabular}{ccccc}
Estimated wave height [cm] & Cell \# & Ruptured cell \# & Total cell \# inhaled per hour & Cell concentration in SW [cells/mL] \\
\hline
$50-100$ & $24$ & $39$ & $7$ & $29.4$\\
$50-100$ & $42$ & $41$ & $9$ & $24.4$ \\
$100-150$ & $81$ & $96$ & $20$ & $4.5$\\
\hline
\end{tabular}

\caption{Airborne cell concentrations during \textit{Ostreopsis} blooms in Saint Jean de Luz (Basque coast, France), summer 2025, with relative estimated wave height and cell concentration in sea water. Cell and ruptured cell number correspond to a $9$~m$^3$ volume of air sampled at $0.3$~L/min. The total (integral and ruptured) number of cells inhaled per hour is calculated considering a rate of volume breath of $1$~m$^3$/h. Cell concentrations in SW were obtained from \cite{paysbasque}.} \label{tab:ecologicaldata}
\end{sidewaystable}

\begin{sidewaystable}
  \centering
     
 \begin{tabular}{lrlccccc}
 & Author and year & Particle type & $r_{p}$~[µm] & $C_{bulk}\,$[mL$^{-1}$] & $R_{b}$\,[mm] & $r_{d}\,$[µm] & $H\,$[cm]\\
\hline
 \circled{1}& Blanchard and Syzdek 1970 \cite{Blanchard1970}  & bacteria & $1.0$ &$[10^3,\;10^6]$ & $[0.1,\;0.5]$ & $[10,\;70] $ & $1.0$ \\
 \circled{2}& Bezdek and Carlucci 1972 \cite{Bezdek1972}& bacteria &  $1.0$ &$[10^4,\;10^6]$ & $[0.1,\;0.7]$ & $[10,\;90]$ & $2.0$\\
 \circled{3}& Blanchard and Syzdek 1972 \cite{Blanchard1972} & bacteria & $1.0$ & $[10^3,\;10^6]$ &  $[0.1,\;0.35]$ & $[10,\;50]$ & $[1.0,\;2]$\\
 \circled{4}& Sakai et al. 1987 \cite{Sakai1987} & microplastics & $0.25$ & $[0.15,\;1.5] \times 10^7$ & $[0.6,2.12]$ & $[100,\;450]$ & $10$\\
 \circled{5}& Ji et al. 2022 \cite{Ji2022} & microplastics & $[0.25,\;1.0]$ & $2.3\times10^6 $ & $[0.19,1.09]$ & $[20,\;410]$ & $[0.5,\;20]$ \\
 \circled{6}& Dubitsky et al. 2023 \cite{Dubitsky2023} & microplastics & $7.5$ or $15$ & $10^4$ or $10^5$ & $[0.25,0.17]$ & $[10,\;120]$ & $5.5$ \\
 \circled{7}& Shaw et al. 2023 \cite{Shaw2023} & microplastics & $[5.0,\;140]$ & $[1.6\times 10^5 ,\; 1.0\times 10^2]$ & $[0.9,1.8]$ &$[160,\;380]$&$1.5$\\
  CR & Present study & \textit{C. reinhardtii} & $4.0\pm 0.8$ & $[3\times 10^2,\; 6\times 10^6]$ & $[0.3,\;2.5]$ & $[25,\;465]$ & $[1.5,\; 25]$\\
  OO & Present study & \textit{O.} cf. \textit{ovata} & $17.6\pm 2.8$ &  $[7\times10^4,\;6\times 10^6]$& $[0.8,\;3.5]$ & $[80,\;375]$& $[1.5,\;2.0]$\\
  LP & Present study & \textit{L. polyedra} & $16.4\pm 2.4$ &  $1.9\times10^4$ & $[1.2,\;2.0]$ & $[195,\;350]$& $[1.5,\;2.0]$\\
  PR & Present study & \textit{Proteomonas} sp. & $2.61\pm 0.21$ &  $1.2\times10^5$ & $[1.3,\;1.3]$ & $[155,\;215]$ & $[1.5,\;2.0]$\\
\hline
\end{tabular}
 
 \caption{Comparison of studies on single bubble bursting in particles and bacteria-laden cultures: particle size $r_p$, bulk concentration $C_{bulk}$, bubble $R_b$ and first drop $r_d$ radii, release distance $H$.}
 \label{tab:litterature}
\end{sidewaystable}

\end{document}